\documentclass[10pt,conference]{IEEEtran}
\usepackage{amsmath}
\usepackage{graphicx}
\usepackage{url}
\usepackage{subcaption}
\usepackage{setspace}
\usepackage{listings}
\usepackage{minted}
\usepackage[table]{xcolor}
\usepackage{tabularx}
\usepackage{threeparttable}
\usepackage{algorithm}
\usepackage{algorithmic}
\usepackage{enumitem}

\usepackage{lipsum}
\usepackage[normalem]{ulem}
\usepackage{xcolor}
\usepackage[htt]{hyphenat}
\usepackage[hidelinks]{hyperref}

\definecolor{codegray}{rgb}{0.92,0.92,0.92}
\setminted{
  fontsize=\small,
  linenos,
  frame=single,
  framesep=2mm,
  tabsize=2,
  breaklines,
  baselinestretch=1.0
}

\newif\ifrevisions
\revisionsfalse

\newcommand{\ourtool}{HINT}

\ifrevisions
\newcommand{\update}[1]{{\color{violet}#1}}
\newcommand{\remove}[1]{{\color{orange}\sout{#1}}}

\newcommand{\simplify}[1]{{\color{blue}\sout{#1}}}
\newcommand{\rjupdate}[1]{{\color{blue}#1}}
\newcommand{\issue}[1]{{\color{red}#1}}
\else
\newcommand{\update}[1]{#1}
\newcommand{\remove}[1]{}

\newcommand{\simplify}[1]{}
\newcommand{\rjupdate}[1]{#1}
\newcommand{\issue}[1]{}
\fi
\providecommand{\Description}[1]{}

\usepackage{tcolorbox}
\tcbuselibrary{skins,breakable}
\newtcolorbox{answerbox}[1]{
  enhanced,
  colback=black!4,
  colframe=black!65,
  boxrule=0.9pt,
  arc=3pt,
  outer arc=3pt,
  left=5pt,
  right=5pt,
  top=4pt,
  bottom=4pt,
  before skip=2pt,
  after skip=4pt,
  fontupper=\small,
  before upper={\textbf{Answer to #1: }\ignorespaces}
}
\usepackage{booktabs}
\usepackage{multirow}
\usepackage{textcomp}
\AtBeginDocument{%
  }

\begin{document}

\title{Hierarchical Fault Localization for Autonomous Driving Systems with Hypothesis Validation and Intent Analysis}

\author{%
\IEEEauthorblockN{Rui~Zheng}%
\IEEEauthorblockA{\textit{KLSS, ISCAS}\\
\textit{Univ. of Chinese}\\
\textit{Academy of Sciences}\\
Beijing, China\\
\texttt{zhengrui@ios.ac.cn}}
\and
\IEEEauthorblockN{Changwen~Li}%
\IEEEauthorblockA{\textit{KLSS, ISCAS}\\
\textit{Univ. of Chinese}\\
\textit{Academy of Sciences}\\
Beijing, China\\
\texttt{licw@ios.ac.cn}}
\and
\IEEEauthorblockN{Yi~Ji}%
\IEEEauthorblockA{\textit{KLSS, ISCAS}\\
\textit{Univ. of Chinese}\\
\textit{Academy of Sciences}\\
Beijing, China\\
\texttt{yeechi18@qq.com}}
\and
\IEEEauthorblockN{Rongjie~Yan}%
\IEEEauthorblockA{\textit{KLSS, ISCAS}\\
\textit{Univ. of Chinese}\\
\textit{Academy of Sciences}\\
Beijing, China\\
\texttt{yrj@ios.ac.cn}}}
\maketitle
\begin{abstract}

Comprehensive testing is essential for the safety and reliability of Autonomous Driving Systems (ADS). Existing techniques can detect system-level failures or attribute them to coarse-grained modules, but they often fall short of localizing the root cause in source code. As a result, debugging remains labor-intensive, requiring developers to connect behavioral violations with complex implementation logic. To address this gap, we present \ourtool{}, a two-phase framework for hierarchical ADS fault localization based on hypothesis validation and intent analysis. In Phase I, \ourtool{} transforms \rjupdate{failure-triggering} execution recordings into multi-modal abstractions and uses causal reasoning to identify the responsible module. In Phase II, it reconstructs design-side intent and implementation-side behavior, then localizes suspicious code through reliability-aware consistency checking, without costly re-simulation. We evaluate \ourtool{} on Apollo across diverse failure modes and modules. The results show that \ourtool{} achieves the strongest overall performance across module-level diagnosis and code-level localization metrics, with 77.8\% end-to-end Class@5 accuracy on real-world bugs.
\end{abstract}

\section{Introduction}
Autonomous Driving Systems (ADS) show great promise in enhancing transportation efficiency and urban mobility. As they are highly safety-critical systems, extensive testing is essential for achieving strong trustworthiness guarantees. Simulation-based testing in high-fidelity environments has become the dominant evaluation method, leveraging its inherent advantages in repeatability, controllability, and cost-effectiveness before real-world deployment.

While diverse methodologies exist for generating simulation tests to evaluate safety and coverage~\cite{lou2022testing,tang2023survey,ding2023survey}, research into the subsequent debugging process remains disproportionately scarce~\cite{Shaw2026ADSdebugging}. 
Though recent attempts have advanced to module-level diagnosis for multi-module ADS, isolating the exact code-level fault remains difficult due to the intricate design and extensive codebase of individual modules. For example, 
recent large-scale empirical studies on platforms like Apollo~\cite{apollo2019} and Autoware~\cite{AutowareUniverse2026} reveal that the {Planning} module is the primary source of failures, involving over five hundred semantic bug-fix patterns, largely driven by intricate sub-modules like path and velocity optimization~\cite{chen2025comprehensive}. Furthermore, these faults rarely manifest as explicit crashes~\cite{jiang2025leveraging}. 

Three key challenges impede code-level fault localization in ADS. 
First, ADS failures often manifest as behavioral deviations rather than code-level exceptions or error messages, making it difficult to connect observed failures with responsible implementation logic. Second, module-level fault reports are insufficient due to the combinatorial explosion of execution paths, making manual root-cause isolation impractical. Third, semantic faults in otherwise syntactically correct code can systematically misdirect both automated debugging and manual analysis, yielding unreliable diagnoses.

\begin{figure*}[htbp]
  \centering
  \includegraphics[width=0.72\textwidth]{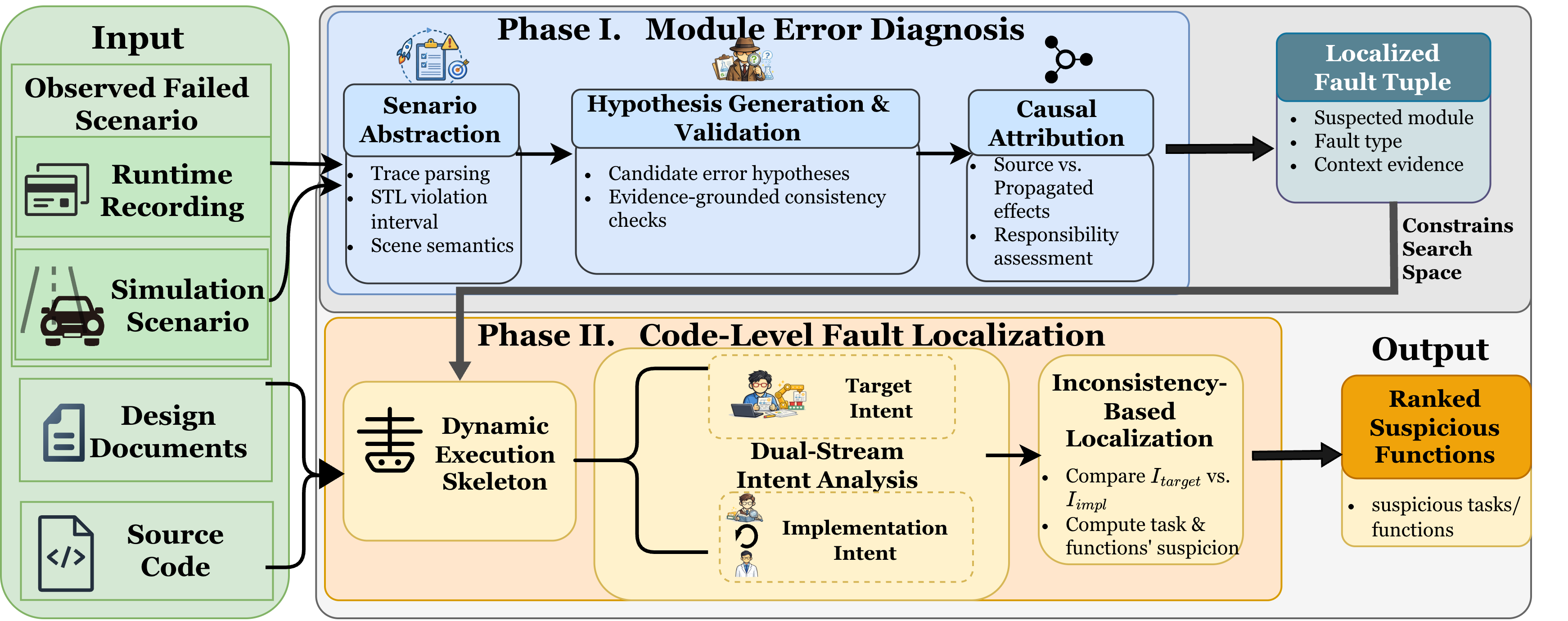}
  \caption{System framework for fault diagnosis and localization}
  \Description{A two-stage process diagram showing Module Error Diagnosis at the top, Environment in the middle, and Code-Level Code Fault Localization Loop at the bottom, coordinated by a Scheduler, ultimately identifying a Fault.}
  \label{fig:framework}
\end{figure*}

To address these challenges, we propose \ourtool{}, a \rjupdate{two-phase} framework for automatically localizing code-level faults of ADS from runtime recordings, ADS documentation, and source code. As shown in Fig.~\ref{fig:framework},
\update{\ourtool{} takes an observed failure trace as input and connects the system-level symptom to the corresponding code-level root cause. 
Currently we 
focus on three failure categories commonly used in scenario-based ADS testing: collisions, traffic-rule violations, and task incompletion.}
The first phase addresses the behavioral-to-code attribution challenge by using LLM-powered agents to analyze synthesized multi-modal evidence from the failure-triggering trace, yielding a candidate faulty module. The second phase addresses search-space explosion by first constructing a scenario-specific Dynamic Execution Skeleton (DES) to prune irrelevant code paths and recover an ordered set of runtime-relevant tasks, and then tackles semantic errors by iteratively checking the consistency between the design intent reconstructed from documentation and the actual implementation, thereby refining localization to specific code segments. \update{This two-phase design is instantiated with ADS-specific evidence, including scenario-based testing results, BEV reconstruction, and STL (Signal Temporal Logic)  failure specifications. The same strategy can be transferred to other modular systems when analogous runtime evidence, contextual information, module dependencies, design artifacts, and source code are available.}

To evaluate the framework's effectiveness and efficiency, we integrated it with the Baidu Apollo platform and the Carla simulator~\cite{dosovitskiy2017carla}. Our benchmark includes execution recordings from public issues and injected faults across ADS modules.
Extensive comparisons with state-of-the-art ADS diagnosis and fault localization techniques demonstrate that our approach achieves high accuracy in pinpointing root causes.

In summary, this paper makes the following  contributions:
\begin{itemize}[leftmargin=*]
\item We present the first framework for automated code-level fault localization in ADS, leveraging system recordings, documentation, and implementation artifacts. \rjupdate{
The framework isolates failure-triggering execution traces to diagnose their code-level root causes. 
}
\item We propose a two-phase hierarchical diagnosis mechanism that combines hypothesis validation with intent analysis to bridge system-level failures and code-level faults.
\item We construct a benchmark of diverse failure scenarios for the Apollo open-source platform, facilitating reproducible evaluation of ADS debugging tools. 
\item We conduct comprehensive experiments to demonstrate the effectiveness and efficiency of \ourtool{} against state-of-the-art fault diagnosis and localization techniques, along with an ablation study on its key design components.
\end{itemize}

\section{Background and Motivating Example}
\subsection{ADS Architecture}
\label{sec:background}

\begin{figure}[htbp]
  \centering
\includegraphics[width=0.8\columnwidth]{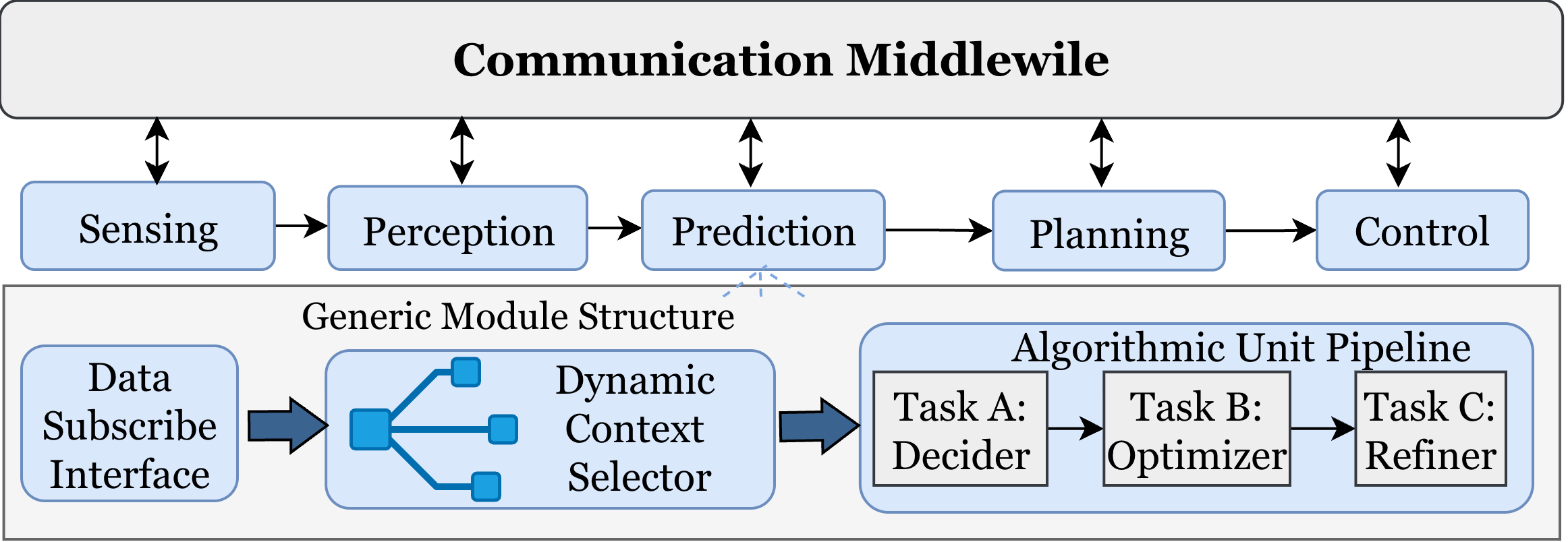}
  \caption{Architecture of Autonomous Driving Systems}
  \label{fig:ads_architecture}
\end{figure}

Modern autonomous driving systems, such as Apollo and Autoware, 
typically adopt a modular architecture centered around a \textit{Communication Middleware}. 
As illustrated in Fig.~\ref{fig:ads_architecture} (middle),
 the system comprises core modules—\texttt{Sensing}, \texttt{Perception}, \texttt{Prediction}, 
 \texttt{Planning}, and \texttt{Control}—that are decoupled via frameworks like 
 Cyber RT~\cite{CyberRT2026} or ROS 2~\cite{Macenski2022ROS2}. 

Crucially, as shown in Fig.~\ref{fig:ads_architecture} (bottom), 
these modules share a generic hierarchical execution model. 
Upon receiving a message trigger, the execution flow is not static but governed 
by a \textit{Dynamic Context Selector} (e.g., Finite State Machines in Apollo or Scene Modules in Autoware). 
This selector determines the active operational context
 (e.g., \textit{Lane Follow} vs. \textit{Intersection}) and dynamically instantiates a specific \textit{Algorithmic Unit Pipeline}. 
This pipeline consists of a sequence of atomic tasks 
(e.g., Deciders and Optimizers) tailored to the current scenario. 

\begin{figure}[t]
  \centering

  \begin{subfigure}[t]{0.8\columnwidth}
    \centering
    \includegraphics[width=0.91\linewidth,height=0.16\textheight,keepaspectratio]{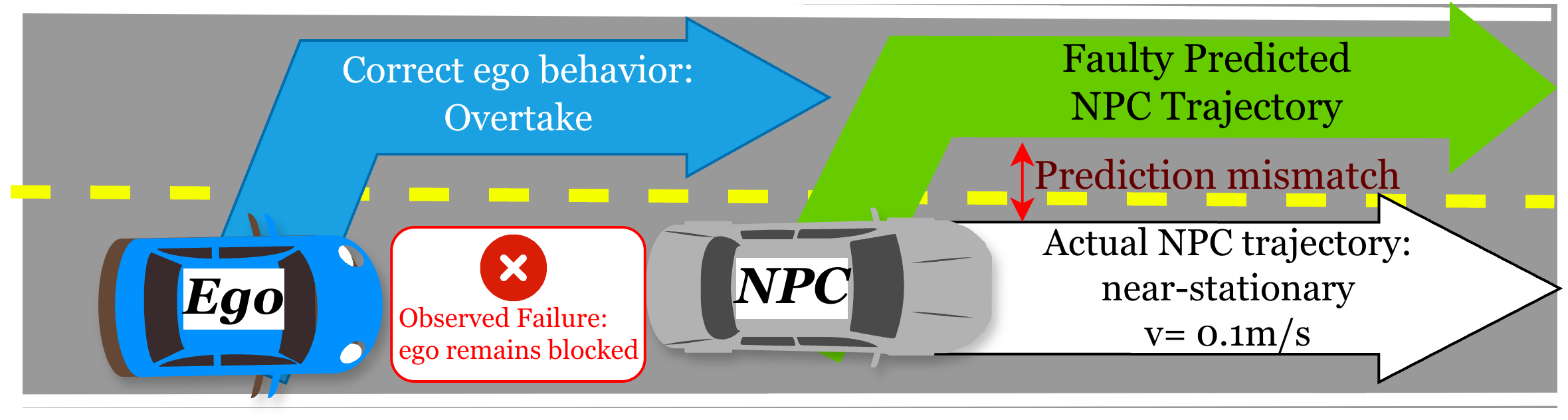}\hspace{4pt}
    \caption{Behavior example}
    \label{fig:motivation_example:a}
  \end{subfigure}

  \begin{subfigure}[t]{0.96\columnwidth}
    \centering
    \includegraphics[width=\linewidth,height=0.16\textheight,keepaspectratio]{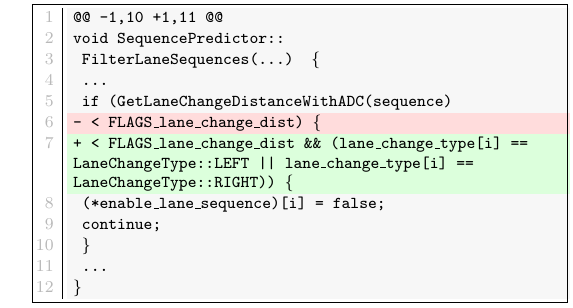}\hspace{13pt}
    \caption{Faulty code before and after the fix}
    \label{fig:motivation_example:b}
  \end{subfigure}

  \caption{Example of a ``failing to overtake'' scenario and the corresponding faulty code fix in PR826~\cite{apollo_pr_826}.}
  \Description{A single-column figure with a behavior example on top and a before-versus-after code comparison below.}
  \label{fig:motivation_example}
\end{figure}

\subsection{Motivating Example}
\label{sec:motivation}

We illustrate the challenges of code-level fault localization in ADS using a real-world bug from the Apollo project (fixed in PR826~\cite{apollo_pr_826}). This bug manifests as a ``failed overtake'' scenario (Fig.~\ref{fig:motivation_example:a}). The blue arrow denotes the expected ego behavior: the ego vehicle should overtake because sufficient space is available. However, the prediction module produces a faulty predicted NPC trajectory, shown in green, indicating that the lead vehicle may obstruct the overtaking maneuver. In contrast, the white arrow shows the actual NPC behavior: the vehicle remains nearly stationary. This mismatch between the predicted and actual NPC trajectories makes the ego vehicle conservatively follow the lead vehicle instead of executing the expected overtake, resulting in the observed failure.

\update{At the system level, the visible symptom is task incompletion. In this scenario, it manifests as a failed overtake: the ego should pass the near-stationary lead vehicle but remains blocked.}
Accurately identifying the cause of this failure requires a method to establish the relationship between the incorrect prediction and the failure to overtake. Moreover, simply knowing the incorrect prediction is still insufficient for localizing the exact faulty code.  For example, in Apollo, the prediction module contains over 29k lines of code and involves a complex workflow with hundreds of functional steps, including geometry computations, map data queries, predicted-trajectory filtering and selection, and machine-learning inferences. The actual bug leading to the
issue is caused by the faulty code shown in Fig.~\ref{fig:motivation_example:b} implementing a filter (function ``\textit{FilterLaneSequences}''). This filter \rjupdate{intends} to restrict the predicted trajectories of surrounding vehicles to keep their lanes when they are close to the ego vehicle, with the goal of avoiding overly cautious behavior by the ego vehicle. However, the erroneous filtering condition removes valid high-confidence trajectories for nearby obstacles, causing subsequent functional steps to operate on an incomplete candidate set and reselect an incorrect trajectory.

\section{Methodology}

\ourtool{} takes an observed failed ADS scenario as input and localizes it to code-level root causes using its runtime recording, simulation scenario, design documentation, and source code. The failure is assumed to exist; STL monitors are used to identify the concrete failure symptom and violation interval for evidence extraction and pruning, rather than to perform general failure detection. Runtime recordings are parsed into time-aligned module traces, simulation scenarios provide physical traffic context, and design documentation and source code are used in Phase II for intent comparison. The workflow consists of two phases:

\begin{itemize}[leftmargin=*]
  \item \textbf{Phase I: Module Error Diagnosis}. 
  This phase abstracts the failure into violation-centered evidence, validates candidate module-error hypotheses, and performs causal attribution to distinguish source faults from propagated or incidental effects. It outputs $E_{localized}=\langle m,b,c\rangle$, where $m$ is the suspected module, $b$ is the violated behavior/fault type, and $c$ is supporting runtime and scenario evidence.

 \item \textbf{Phase II: Code-Level Fault Localization}. 
 Given $E_{localized}$, design documents, and source code, this phase localizes the fault within the diagnosed module. It reconstructs the design-side task intent,
 builds a Dynamic Execution Skeleton (DES) to prune irrelevant code paths under the observed failure, extracts the implementation-side intent
 from runtime-relevant source code, and ranks suspicious tasks/functions by semantic inconsistencies under a shared schema.

\end{itemize}

\subsection{Phase I: Module Error Diagnosis}

Phase I isolates the module error responsible for a system-level failure through evidence-grounded causal analysis. It extracts structured evidence around the violation episode, derives candidate module-level inconsistencies, and validates which candidate best explains the failure rather than merely reflecting propagated, compensatory, or incidental behavior.

\subsubsection{Step 1: Scenario Abstraction}

\begin{figure}[htbp]
  \centering
  \includegraphics[width=0.9\columnwidth]{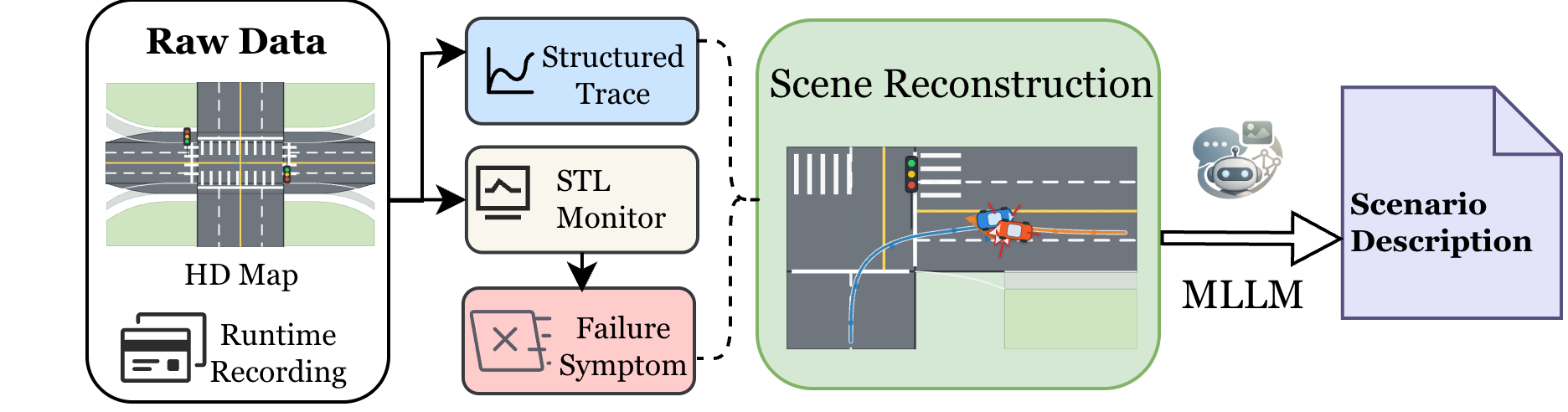}
  \caption{The scenario abstraction pipeline}
  \Description{A flowchart showing the abstraction process from raw data (HD Map and Record) through abstraction (Visualizer and STL Monitor) to fusion (MLLM generating scenario descriptions).}
  \label{fig:abstraction}
\end{figure}

Raw runtime artifacts are noisy and difficult to reason over directly. As shown in Fig.~\ref{fig:abstraction}, \ourtool{} converts them into a violation-centered, multi-modal representation for causal diagnosis.

First, we parse heterogeneous messages in the raw recording
into a structured, time-aligned  execution trace with  synchronized outputs of all modules.  This transformation converts raw logs
into a discrete, queryable sequence, providing the foundational evidence required for subsequent
causal reasoning.

Second, we extract time-series signals (e.g., velocity and distance) from the abstracted trace and employ STL monitors~\cite{maler2004monitoring,rtamt2024,sun2022lawbreaker} to
\rjupdate{support the precise localization of violation intervals from a failure symptom $\mathcal{F}$,
such as collision, traffic-rule violation, or task incompletion. The violation interval $\Delta_v$ captures the specific timeframe within a trace during which the robustness value of the corresponding STL formula remains below zero.}

Third, \ourtool{} reconstructs the traffic scene by projecting the parsed execution trace onto the HD map. The resulting BEV overlays static map elements with ego and surrounding-vehicle trajectories, exposing spatial context such as lane occupancy, relative distances, and motion trends that are implicit in raw logs.

Finally, to unify the topological context from the 
visualizer and logical verdicts from the STL monitor, we employ a 
\textit{Multimodal LLM (MLLM)}. 
\rjupdate{It synthesizes the BEV imagery and quantitative signal data into a \textit{Scenario Description} in natural language. }
 This step effectively translates low-level runtime artifacts into a unified semantic format that helps the Hypothesis Agent organize the scenario context for downstream analysis.

\subsubsection{Step 2: Hypothesis Generation and Validation}

In Step 2, \ourtool{} isolates the faulty module using a dynamic Hypothesis Generation and Validation process (Alg.~\ref{alg:hypothesis_loop1}). The process begins with the Hypothesis Agent generating a set of coarse-grained hypotheses from the observed symptom and the violation interval, while the localized fault tuple is initially left unknown (Lines 1--2).
These hypotheses are constrained by the ADS pipeline and the anomaly type, preventing the divergence associated with open-ended generation.

The diagnosis then proceeds iteratively as long as unresolved hypotheses remain and the current fault entry is not yet sufficiently specific (Line 3). In each iteration, the Hypothesis Agent selects the most promising hypothesis and maps it to an appropriate validation \emph{Skill}, a reusable diagnostic operator that specifies the evidence to inspect and the runtime consistency check to perform for the selected hypothesis. The selected Skill queries the structured execution trace and reconstructed scene to produce a grounded observation. This design allows each suspect to be examined through a validation procedure tailored to its semantic role.

\rjupdate{Validation function (Line 6) is a structured, dependency- and semantics-aware check, not mere anomaly detection.} More specifically, validation checks whether a candidate module's behavior is consistent with runtime evidence. 
\rjupdate{Its output should be 1) explainable by its inputs and recent observations, 2) consistent with the downstream behavior observed in actual execution, and 3) semantically coherent under the current map and obstacle context.}
Different hypotheses are examined through different Skills, each instantiating a corresponding form of behavioral consistency analysis.

Based on the validated observation, the Hypothesis Agent then performs causal responsibility analysis (Line 7) to determine whether the inconsistency is best explained as a source fault, a propagated effect, a compensatory response, or an incidental anomaly. This judgment considers the timing of the inconsistency, the module's role in the ADS pipeline, and the explanatory power of the evidence. As a result, \ourtool{} identifies the module whose abnormal behavior most plausibly explains the observed system-level failure, rather than simply reporting the first abnormal one.

If the evidence contradicts the current hypothesis, the suspect is pruned from the candidate set (Lines~8--9). Otherwise, the localized fault tuple is updated, and the hypothesis is refined as needed (Lines~11--13). The loop terminates when the candidate set is exhausted or the diagnosis becomes sufficiently specific, following the loop condition in Line~3. It returns a localized fault tuple that records the suspected module, failure type, execution context, and supporting evidence for Phase~II.

\begin{algorithm}[t]
\small
\caption{Hypothesis Generation and Validation}
\label{alg:hypothesis_loop1}
\begin{algorithmic}[1]
\REQUIRE Failure Symptom $\mathcal{F}$, Violation Interval $\Delta_v$, Agent Skills $\mathcal{S}$
\ENSURE Localized Fault Tuple $E_{localized}$
\STATE $\mathcal{H} \leftarrow \text{InitializeHypotheses}(\mathcal{F}, \Delta_v)$
\STATE $E_{localized} \leftarrow \text{Unknown}$
\WHILE{$\mathcal{H} \neq \emptyset$ \textbf{and not} $\text{IsSpecific}(E_{localized})$}
\STATE $h \leftarrow \text{SelectTopHypothesis}(\mathcal{H})$
\STATE $s \leftarrow \text{SelectSkill}(\mathcal{S}, h)$
\STATE $O \leftarrow \text{ValidateHypothesis}(s, h)$ \COMMENT{trace/scene evidence}
\STATE $r \leftarrow \text{AssessResponsibility}(h, O, \Delta_v)$ \COMMENT{fault attribution}
\IF{$r = \text{Rejected}$}
\STATE $\mathcal{H} \leftarrow \mathcal{H} \setminus {h}$ \COMMENT{Prune}
\ELSE
\STATE $E_{localized} \leftarrow \text{UpdateFaultEntry}(h, O, r)$
\IF{$\text{NeedsRefinement}(h, O, r)$}
\STATE $\mathcal{H} \leftarrow (\mathcal{H} \setminus {h}) \cup \text{RefineHypothesis}(h, O)$
\ENDIF
\ENDIF
\ENDWHILE
\RETURN $E_{localized}$
\end{algorithmic}
\end{algorithm}

\subsection{Phase II: Code-Level Fault Localization}

\begin{figure*}[t]
  \centering
\includegraphics[width=0.85\textwidth]{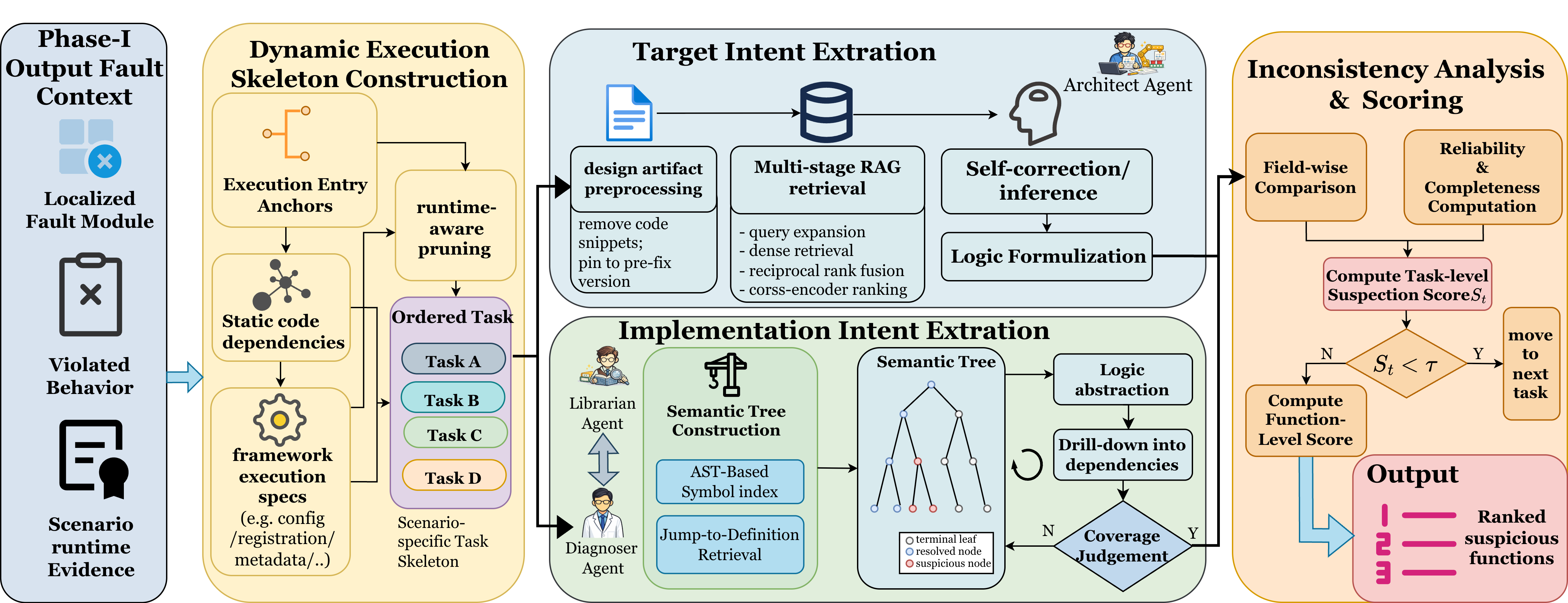}
  \caption{Overview of Phase II: Dual-Stream Intent Analysis for code-level fault localization.}
  \Description{A process diagram showing Intent Extraction (top path) and Implementation Extraction (bottom path) converging at Alignment to identify the Root Cause from a Module Error input.}
  \label{fig:phase2_arch}
\end{figure*}

To address search space explosion and semantic-error challenges in code-level fault localization, we propose a \textit{Dual-Stream Intent Inconsistency Analysis} process (Fig.~\ref{fig:phase2_arch}).

As mentioned in Section~\ref{sec:background}, ADS execution can be modeled as a sequence of \textit{tasks}, where each task denotes a coherent design/implementation unit, such as an Apollo decider, optimizer, predictor, or trajectory-filtering component. Given the Phase-I output, design documentation, and source code, Phase II reconstructs the \textit{Dynamic Execution Skeleton} (\textit{DES}), a scenario-specific ordered sequence of runtime-relevant tasks that approximates the execution path under the observed failure. Here, \textit{task} does not refer to the ego vehicle's driving maneuver or an \rjupdate{operation} performed by \rjupdate{the agents in the framework}.

Rather than constructing a global program graph over the entire codebase, \ourtool{} focuses on the code region around the localized faulty module. It first identifies \textit{execution entry anchors}, such as message callbacks, periodic execution handlers, or framework-registered dispatch points, that can initiate the relevant runtime logic, then recovers their static code dependencies and uses framework-level execution specifications (e.g., configuration files, registration metadata, or pipeline descriptors) as structural hints for task ordering and activation. It then incorporates runtime-relevant scenario evidence from Phase I to perform runtime-aware pruning, removing infeasible branches and retaining only tasks and bindings compatible with the observed failure. The resulting skeleton provides an ordered, scenario-specific task list and a runtime-aware boundary for subsequent code retrieval. Tasks are then analyzed in this order, with later candidates considered only if earlier ones are rejected. For each selected task, Phase II performs two streams of analysis:

 \begin{enumerate}[leftmargin=*]
  \item \textbf{Target intent Extraction:}
  Reconstructing the \textit{Target Intent} (${I}_{target}$) strictly from design artifacts.

  \item \textbf{Implementation Intent Extraction:}
    Reconstructing the Implementation Intent ($I_{impl}$) strictly from the source code, through an information-gain-driven semantic tree construction that selectively expands only behaviorally relevant function calls while folding infrastructure code.

\end{enumerate}

The two streams converge only at the \textit{Inconsistency Analysis} stage, where ${I}_{impl}$ is systematically compared against ${I}_{target}$ across multiple semantic dimensions to quantify their discrepancy. If the inconsistency score exceeds the acceptance threshold, the root cause is localized, and the score is distributed onto the semantic tree's function nodes to pinpoint the responsible code region. Otherwise, the system proceeds to the next task in the skeleton.

To make the two streams directly comparable, we adopt a unified intent schema for Phase II. 
\update{The schema is fixed by the framework rather than learned from each scenario. It is used to normalize both design-side evidence and implementation-side behavior so that comparison is performed field by field under the same semantic structure.} 
For each selected task, an intent is represented as a quadruple
$I = \langle Goal, Guard, Safety, Assumption \rangle$. 
Here, \textit{Goal} denotes the functional objective that the task is expected to achieve; \textit{Guard} specifies the activation conditions under which the task behavior becomes relevant; \textit{Safety} captures the mandatory constraints that the task output must satisfy; and \textit{Assumption} records the environmental premises on which the task logic relies. Each field is instantiated as a structured natural-language semantic slot rather than a symbolic formula, allowing evidence from heterogeneous sources to be normalized into a shared representation. 

\subsubsection{Target Intent Extraction}

This stage reconstructs the design-side intent $I_{target}$ for the selected task strictly from design artifacts, operating in a code-blind mode to ensure that the resulting representation is not biased by the observed implementation. The goal is to recover what the task is intended to do according to the design knowledge available to developers, rather than what the current code happens to implement. The extraction proceeds in two steps: \emph{Information Retrieval \& Inference} and \emph{Logic Formalization}, both carried out by an \emph{Architect} agent, which serves as the design-side reasoning component in Phase II.

\smallskip
\textit{Information Retrieval \& Inference.} The purpose of this step is to identify the design artifacts most relevant to the selected task and recover from them the design knowledge needed for intent reconstruction. To this end, the Architect first organizes heterogeneous design artifacts, including requirements documentation, interface specifications, protocol descriptions, and technical reports, into a searchable knowledge base. 

 \rjupdate{To recover the design knowledge most relevant to the selected task under the observed failure scenario\cite{cormack2009reciprocal,nogueira2019passage}, 
 the Architect 
  performs retrieval through a \textit{Multi-stage Retrieval-Augmented Generation (RAG)}   \cite{lewis2020retrieval} pipeline.}  Starting from the fault context and the selected task, it formulates multiple semantic queries, retrieves candidate passages from the design corpus, and then reranks them to retain the most relevant evidence for intent reconstruction.

In practice, industrial documentation is often incomplete, fragmented, or outdated.
As a result, retrieval alone may be insufficient for reconstructing a coherent design-side intent,
when it does not provide enough grounded evidence to instantiate the key fields of $I_{target}$ with reasonable consistency. 
In such cases, the Architect triggers a \textit{self-correction/inference} step that supplements the documented intent with plausible implicit constraints derived from domain knowledge and the task's functional role. This step is not intended to invent new behavior, but to recover design semantics that are strongly implied yet not explicitly stated in the available artifacts.

\textit{Logic Formalization.}
\rjupdate{
In this step, the Architect synthesizes the retrieved specifications and inferred constraints into a coherent task-level design intent.} Each field is instantiated in structured natural language, preserving semantic richness while maintaining cross-source comparability.

When retrieved design evidence is too abstract for direct comparison, the Architect performs \textit{Constraint Instantiation} to concretize under-specified parts of the design-side intent. By applying counterfactual reasoning to the observed failure, it infers latent numerical or logical bounds necessary to operationalize the intended behavior, such as embedding $\theta_{stop} \leq 12.3~m$ into a relevant \textit{guard} condition. 
\rjupdate{When fields cannot be reliably grounded through documentation or domain-informed inference, the Architect flags them as \textit{under-specified} to avoid introducing spurious content.}

\subsubsection{Implementation Intent Extraction}
\label{sec:implementation_extraction}

This stage reconstructs the implementation-side intent $I_{impl}$ for the selected task strictly from the source code, operating in a specification-blind mode to ensure that the resulting representation captures only what the code actually does, independent of what it was intended to do. Two agents facilitate the extraction process: the \textit{Diagnoser} performs the implementation-side abstraction,
while the \textit{Librarian} manages an AST-based symbol index for efficient 
\textit{Jump-to-Definition} retrieval (resolving a symbol reference or call site to the corresponding function or method definition). Static references are resolved via the index, while dynamic symbol bindings are handled through the DES established at the start of Phase II.
Building on this infrastructure, \ourtool{} performs \textit{Implementation Intent Extraction} through three tightly coupled components: \textit{Semantic Tree construction}, which organizes behaviorally relevant code entities around the selected task; \textit{Logic Abstraction}, which summarizes the recovered code semantics under the shared intent schema; and \textit{Convergence Judgment}, which determines when recursive expansion is sufficient for a self-contained root-level characterization.

\textit{Semantic Tree Construction:}
A \emph{Semantic Tree} is a hierarchical representation of behaviorally relevant implementation logic rooted at the selected task entry point. Unlike a raw call graph,
it selectively includes only callees, predicates, helper routines, and delegated checks whose semantics may affect the task-level characterization. Each node corresponds to a code entity or logic fragment together with its local semantic contribution, while
edges capture semantic dependence induced by control flow, data flow, or delegated decision logic. The subsequent implementation-side analysis is then carried out over this tree: the Diagnoser iteratively expands, abstracts, and checks the Semantic Tree until the extracted task intent converges.

\textit{Logic Abstraction:} Starting from the selected task entry point as the root of the Semantic Tree, the Diagnoser abstracts the AST-parsed code into $I_{impl}$.
Because the semantics of a task are rarely self-contained within a single function body, the Diagnoser performs \textit{Drill-Down} whenever the current abstraction contains an opaque dependency, such as a delegated safety check or an unresolved helper function. Built on top of \textit{Jump-to-Definition}, \textit{Drill-Down} follows the relevant dependency to its defining implementation, expands the Semantic Tree under the runtime-aware bindings established by the DES, and propagates the newly exposed behavior back into the current abstraction. To prevent the tree from expanding indefinitely while still preserving all behaviorally relevant semantics, this recursive process is coupled with a convergence criterion that determines when the current abstraction is already sufficient for task-level reasoning.

\textit{Convergence Judgment:}
Let $T^{i}$ be the Semantic Tree after $i$-th iteration, and let $F^{i} \subseteq \mathrm{leaves}(T^{i})$ be the set of unresolved leaf nodes, each containing at least one opaque dependency that may still affect the task-level characterization. At each iteration, the Diagnoser selects one candidate $a \in F^{i}$ for trial expansion: $\mathrm{Expand}(T^{i}, a)$,
where 
$\mathrm{Expand}$ resolves the dependency, retrieves the relevant implementation, and incorporates any newly exposed semantics into the tree. The expansion is accepted only if it introduces a substantive update to the current task-level intent; otherwise, the candidate is sealed as a terminal leaf and no longer considered for further expansion. In this way, each iteration either expands the tree with semantically useful content or explicitly marks a dependency as irrelevant to the task-level characterization. Note that the exact traversal order is not essential: unresolved leaves may be examined in different orders, but convergence is always determined by whether the expansion changes the current task-level implementation intent.

The tree construction converges when no unresolved leaf can further refine the current implementation intent:
\[
\forall a \in F^{i}:\;
I_{impl}(\mathrm{Expand}(T^{i}, a)) = I_{impl}(T^{i}),
\]
where $I_{impl}(T)$ denotes the task-level implementation intent extracted from
$T$. At this point, every dependency that could materially affect the selected task intent has either been resolved into the tree or determined to be semantically inert.

Because the intent fields are represented in structured natural language rather than symbolic normal forms, the equality test above is operationalized as a \emph{semantic non-update} judgment: the Diagnoser retains an expansion only if the newly exposed content materially refines the current task-level characterization, rather than merely adding local implementation detail. The final output of this process is a converged implementation-side intent $I_{impl}$ for the selected task, which serves as the code-grounded counterpart to $I_{target}$ for subsequent inconsistency analysis.

\subsubsection{Inconsistency Analysis}
\label{sec:inconsistency_analysis}

This stage assigns each selected task a suspicion score by jointly considering semantic discrepancy and evidence reliability. A task is ranked as highly suspicious only if its design-side and implementation-side intents exhibit a substantive mismatch, and that mismatch is supported by trustworthy evidence. A mismatch alone is not sufficient: if the design-side intent is weakly grounded or the implementation-side abstraction remains incomplete, the apparent discrepancy may reflect uncertainty rather than a true fault signal.

Accordingly, \ourtool{} performs field-wise semantic comparison between $I_{target}$ and $I_{impl}$ and calibrates the resulting discrepancy signals by reliability factors. These factors reflect the grounding of the retrieved design evidence and the completeness of the current Semantic Tree expansion. The final task-level suspicion score therefore captures both the magnitude of the intent divergence and the confidence in that divergence. For suspicious tasks, \ourtool{} propagates the task-level score to functions according to how directly each function contributes to the observed inconsistency.

\textit{Field-wise Comparison:}~For each field 
$d\in \mathcal{D}=\{\textit{Goal}$, $\textit{Guard}$,$\textit{Safety},\textit{Assumption}\}$, \ourtool{} compares $I_{target}[d]$ and $I_{impl}[d]$ and assigns a discrepancy score
$\delta_d(t)\in[0,1]$ for task $t$. Each field has a fixed weight
$\lambda_d\in(0,1]$, \rjupdate{where safety-related mismatches are weighted higher than auxiliary assumption drift.}
The inconsistency strength of task $t$ is then computed as:
\[
V(t)=\max_{d\in\mathcal{D}}\bigl[\lambda_d\cdot\delta_d(t)\bigr].
\]

We use $\max$ because one high-confidence inconsistency in a critical field is sufficient to implicate a task; when all fields are judged consistent under the comparison criteria, all discrepancy scores are zero and thus $V(t)=0$.

\textit{Evidence Reliability:}
To calibrate the discrepancy signal, \ourtool{} estimates reliability from both the design and implementation sides. For the design side, we define
$R_g(t)=\alpha\cdot\bar{r}(t)$, where $\bar{r}(t)$ denotes the retrieval support quality of the design evidence for task $t$, and $\alpha\in(0,1]$ discounts intent fields inferred from incomplete documentation rather than directly supported by retrieved artifacts. For the implementation side, $R_c(t)$ measures code-extraction completeness as the ratio of successfully retrieved and analyzed semantic dependencies to all dependencies encountered during the Semantic Tree expansion of task $t$. Thus, when the design intent is weakly supported or the code exploration is incomplete, the observed inconsistency has less influence on the final suspicion score.

\textit{Fault Suspicion Score:}
Given the inconsistency strength $V(t)$ and reliability factors $R_g(t)$ and $R_c(t)$, \ourtool{} computes a task-level suspicion score:
\[
S(t) = p_0 + (V(t)-p_0)\cdot R_g(t)\cdot R_c(t).
\]
where $p_0=\frac{1}{N}$ is the uniform baseline over $N$ candidate tasks. This formulation lets the discrepancy signal raise a task’s suspicion above the uninformed baseline only when both the design-side and implementation-side evidence are reliable; when either side is weak, the score falls back toward $p_0$. \ourtool{} then compares $S(t)$ with a pruning threshold $\tau$: tasks with $S(t)<\tau$ are discarded, while the remaining tasks are retained for function-level localization. Thus, $\tau$ affects search efficiency by controlling how aggressively low-plausibility tasks are pruned, without changing the score definition.

\textit{Function-level Suspicion Score:}
For tasks whose suspicion scores pass the pruning threshold, \ourtool{} propagates suspicion to functions in the semantic subgraph $G(t)$, since a suspicious task may still involve several interacting functions. We use function-level rather than statement-level ranking because ADS logic faults often span related predicates, branches, and helper calls along the same execution path. Each function $f$ receives a relevance weight $\omega(f,t)\in(0,1]$ based on how directly the inconsistency implicates it:
\[
\hat{S}(f,t)=S(t)\cdot \frac{\omega(f,t)}{\sum_{f'\in G(t)} \omega(f',t)}.
\]
If a function appears in multiple tasks, \ourtool{} keeps maximum score as the final ranking value.

\section{Implementation}
\label{sec:implementation}

We implement \ourtool{} as an agentic fault-localization prototype for Baidu Apollo~\cite{apollo2019}, with platform-specific logic isolated behind adapters. LangGraph~\cite{langgraph2024} orchestrates three components: a runtime-evidence processor, a diagnostic-skill library, and a dual-stream intent-analysis engine. The processor parses Apollo records into time-aligned traces, identifies violation intervals with STL monitors, and reconstructs scene evidence from simulator states and HD maps. Each Skill is implemented as a reusable Phase-I validation operator that declares its applicable hypothesis type, required runtime evidence, consistency checks, output schema, and supporting scripts/templates, grounding hypothesis validation in executable evidence queries rather than open-ended LLM reasoning.

The functions in Alg.~\ref{alg:hypothesis_loop1} are implemented as function-specific LLM-agent calls with constrained prompts and lightweight post-processing. Phase I uses separate calls for hypothesis initialization, ranking, Skill selection, validation summarization, and responsibility assessment, then normalizes the validated result into $E_{localized}=\langle m,b,c\rangle$. Phase II uses a shared $\langle Goal, Guard, Safety, Assumption\rangle$ schema for target and implementation intents: target-intent calls require retrieved design evidence or an under-specified marker, while implementation-intent calls are restricted to source-code evidence from the code index and DES. Lightweight subtasks use Claude Haiku~\cite{anthropic_claude}; responsibility assessment, intent abstraction, and inconsistency judgment use Claude Sonnet. All LLM calls use fixed model identifiers, fixed decoding settings, and bounded retries; prompts, Skill templates, settings, retry records, and per-case outputs are released in the replication package~\cite{hint_replication_package}. Design documents are indexed with BAAI/bge-m3~\cite{flagembedding2023} and ChromaDB~\cite{chromadb2024}; C++ code navigation uses a Tree-sitter~\cite{treesitter2024} AST symbol index.

\section{Evaluation}

In this section, we evaluate the effectiveness of \ourtool{} through comprehensive experiments. Our evaluation aims to answer the following research questions:

\begin{itemize}[leftmargin=*]

\item \noindent\textbf{RQ1.} How effective is \ourtool{} at module-level fault diagnosis?

\item \noindent\textbf{RQ2.} How effective is \ourtool{} at code-level fault localization?

\item \noindent\textbf{RQ3.} How much do the key components of \ourtool{} contribute to accuracy and efficiency?

\item \noindent\textbf{RQ4.} How effective is \ourtool{} as an end-to-end fault localization pipeline, and how robust is it across different LLM backbones?
\end{itemize}

\subsection{Experimental Setup}

We establish \ourtool{} a high-fidelity co-simulation environment using Apollo 9.0~\cite{apollo2019} and Carla 0.9.14~\cite{dosovitskiy2017carla} on Ubuntu 22.04, with an NVIDIA RTX 4090 GPU. Unless otherwise stated, all experiments use the Apollo implementation described in Section \ref{sec:implementation}. We set the pruning threshold $\tau$ to 0.3 for all experiments. This threshold is applied only after task-level suspicion scores are computed and therefore mainly controls how aggressively low-suspicion candidates are filtered rather than changing the score definition itself.

\subsubsection{Benchmark Construction}
We construct two complementary datasets to evaluate \ourtool{}, as summarized in Table~\ref{tab:benchmark}. \update{For each benchmark case, we treat the latest ADS version as the reference system and construct a faulty variant either by replacing a function with its counterpart from an earlier version or by introducing a synthetic semantic fault. We retain a test case only when the reference ADS passes and the faulty ADS fails under the corresponding oracle. Each retained case is executed at least three times to reduce nondeterministic pass/fail outcomes, and two experienced developers inspect runtime logs to confirm that the faulty code triggers the unexpected behavior. The benchmark covers three STL-based oracle categories: collisions, traffic-rule violations, and task-completion failures. In our benchmark, the delay between the fault-triggering behavior and the observed failure is up to 5 seconds; longer-delayed failures are beyond the scope of the current evaluation.}

\update{
\subsubsection{Datasets}  The injected-fault dataset, $D_{inject}$, contains semantic faults introduced into Control, Planning, and Prediction, including inverted guard conditions, missing safety checks, and incorrect threshold values. The real-world dataset, $D_{real}$, consists of historical Apollo bugs reproduced from public commits and pull requests. 
\rjupdate{These real-world cases evaluate \ourtool{} under realistic fault patterns. While their sample size is inherently constrained, this reflects the natural scarcity of reproducible, safety- or performance-critical public bugs compared to the high prevalence of routine refactoring, documentation, or maintenance changes.}

\subsubsection{Baselines and Metrics}
We compare \ourtool{} with baselines at both stages. For Phase I, we use ACAV~\cite{sun2024acav}, which infers causal events from AV accident recordings, and an Oracle-based baseline following MoDitector~\cite{wang2025moditector}, which flags a module when its output violates a module-specific oracle. For Phase II, we use LocAgent~\cite{chen2025locagent}, AutoFL~\cite{kang2024quantitative}, and Claude Code as a representative repository-level code agent. LocAgent performs graph-guided repository navigation, while AutoFL explores functions and call relations with LLMs. For a fair comparison, all Phase-II baselines receive the same module-level context as \ourtool{}, including the failure description, annotated faulty module, design artifacts, and repository access. We omit spectrum-based fault localization because our setting provides a single failure-triggering trace rather than a passing/failing coverage suite. For end-to-end evaluation, composing Phase-I and Phase-II baselines is not well-defined because existing Phase-I methods do not output the structured evidence required by Phase II. We therefore use approximate generic-agent baselines: Claude Code receives the failure record/STL report, design artifacts, and repository, while Claude Code + Skills additionally receives our diagnostic Skill scripts. Neither baseline uses \ourtool{}'s selected module, DES, or intent-analysis outputs. Claude Code uses the same Claude Sonnet backbone as \ourtool{}'s reasoning-heavy agents.

\rjupdate{For Phase I, we report module-level accuracy. For Phase II, we use \textit{Top-$k$ Accuracy (Acc@$k$)}, which checks whether the ground-truth faulty element appears within the top-$k$ ranked candidates. We report Class@$1/3/5$ and Function@$3/5/10$, matching the class- and function-level rankings produced by \ourtool{}. Function-level results are reported instead of statement-level results because ADS logic faults often span related predicates, branches, and helper calls along the same execution path.} For isolated Phase-II evaluation, all methods are given the same ground-truth faulty module; for end-to-end evaluation, \ourtool{} uses the Phase-I predicted module as input to Phase II.
}

\begin{table}[t]
  \centering
  \caption{Benchmark Dataset Composition}
  \label{tab:benchmark}
  \small
  \begin{tabular}{l|ccc|c}
  \toprule
  \textbf{Dataset} & \textbf{Control} & \textbf{Planning} & \textbf{Prediction} & \textbf{Total} \\
  \midrule
  $D_{inject}$ & 19 & 25 & 19 & 63 \\
  $D_{real}$ & 1 & 5 & 3 & 9 \\
  \midrule
  \textbf{Total} & 20 & 30 & 28 & 72 \\
  \bottomrule
  \end{tabular}
\end{table}

\subsection{RQ1. Module Error Diagnosis}

This research question evaluates the framework's ability to identify the faulty module responsible for a system-level failure. Since no existing ADS fault localization approach directly traces from system-level failures to faulty code, we first evaluate module-level diagnosis independently against two representative baselines.

\begin{figure}[t]
  \centering
  \includegraphics[width=\columnwidth]{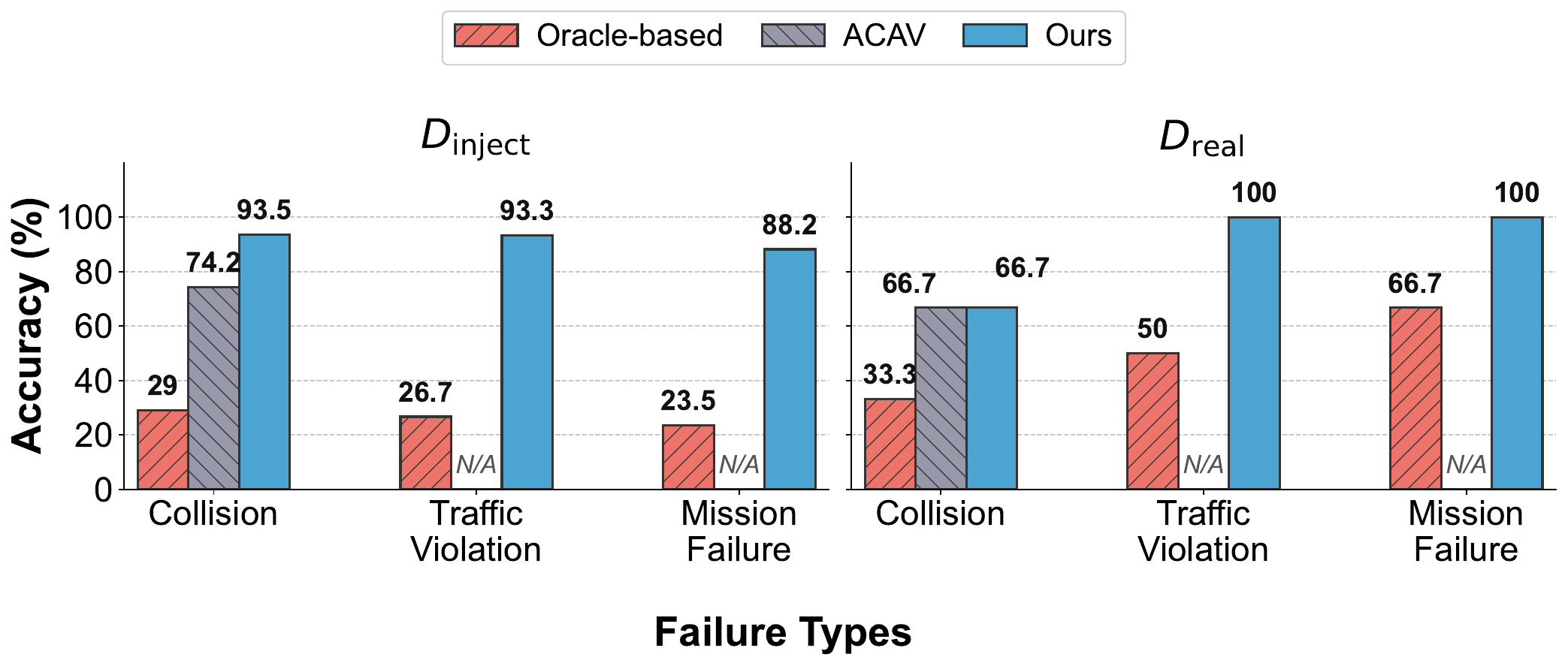}
  \caption{Comparison of Module Diagnosis Accuracy}
  \Description{Bar chart comparing module diagnosis accuracy across Oracle-based, ACAV, and KORAL methods for different failure types.}
  \label{fig:module-diagnosis-accuracy}
\end{figure}

Fig.~\ref{fig:module-diagnosis-accuracy} shows that \ourtool{} substantially outperforms both baselines on $D_{inject}$ and $D_{real}$, indicating that it generalizes beyond collision-centric diagnosis and handles diverse ADS failure types. This improvement mainly comes from validating candidate module-error hypotheses against runtime and scenario evidence, rather than simply flagging abnormal module outputs. Oracle-based diagnosis is sensitive to error propagation and lacks causal attribution: a downstream abnormal output does not necessarily identify the source fault. ACAV remains partially competitive on collisions, but its collision-centered causal model does not cover traffic-rule violations or task-completion failures well.

\begin{figure}[t]
  \centering
  \small
  \begin{subfigure}[t]{0.38\columnwidth}
    \centering
    \includegraphics[width=\linewidth]{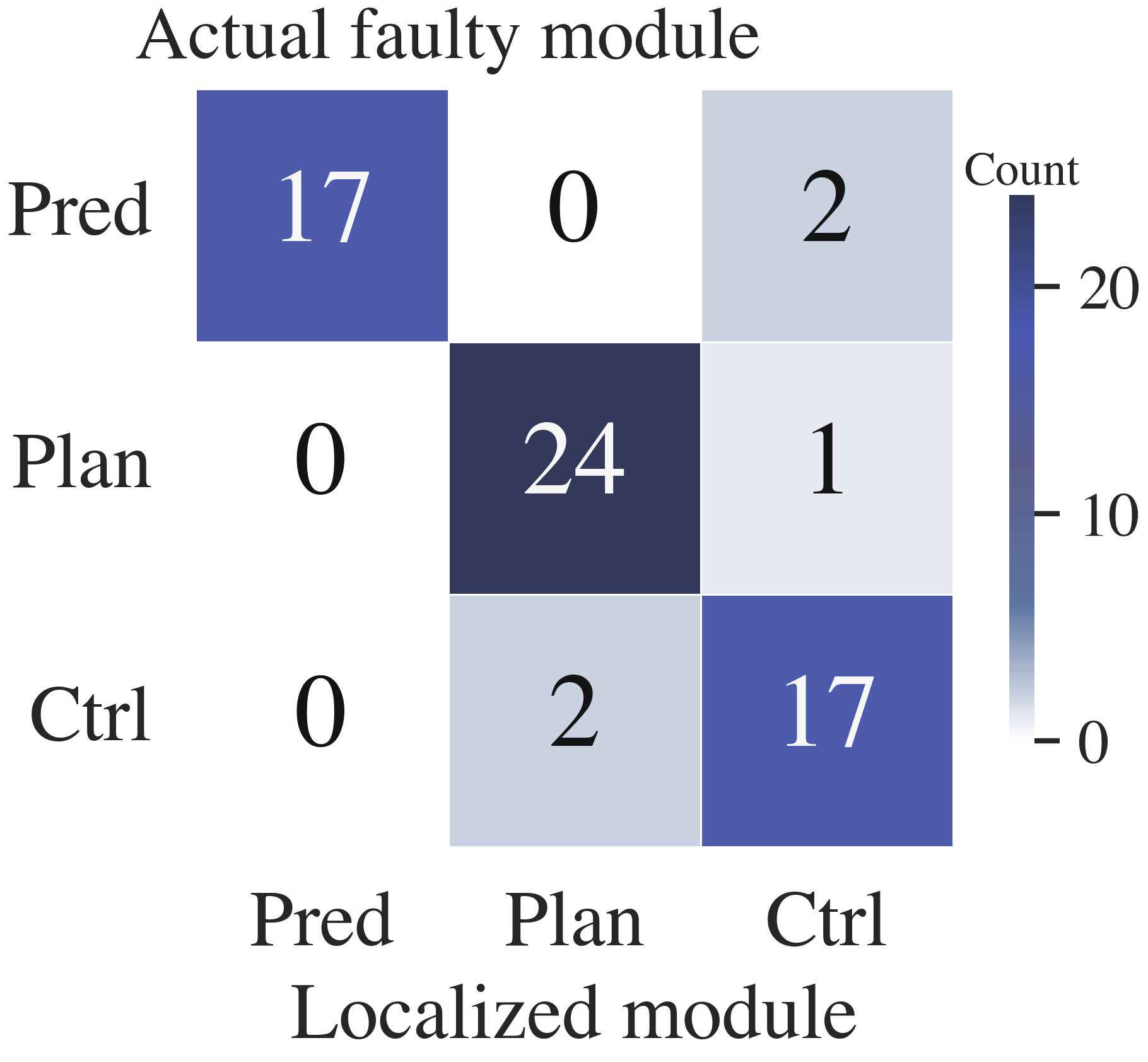}
    \caption{$D_{inject}$}
  \end{subfigure}
  \hspace{0.02\columnwidth}
  \begin{subfigure}[t]{0.38\columnwidth}
    \centering
    \includegraphics[width=\linewidth]{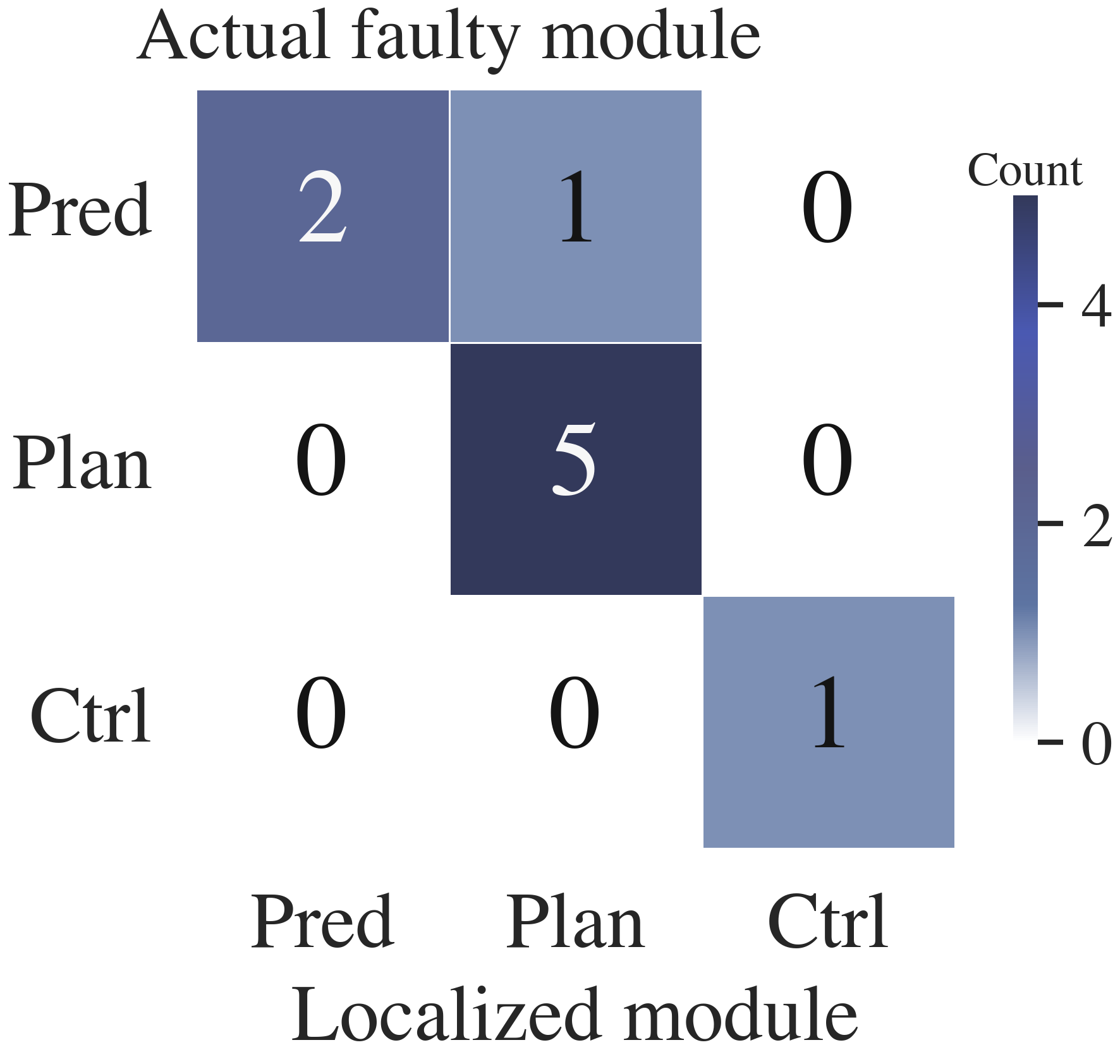}
    \caption{$D_{real}$}
  \end{subfigure}
  \normalsize
  \caption{Confusion matrix of actual vs. localized faulty module
  }
  \Description{Two heatmaps side by side: left labeled Inject for injected faults dataset, right labeled Real for real bugs; rows are actual faulty module and columns are reported module, with count-based color scale.}
  \label{fig:module_confusion}
\end{figure}

Fig.~\ref{fig:module_confusion} further shows that diagonal entries dominate on both datasets and \ourtool{}'s module assignments do not collapse to one frequent module. The few off-diagonal errors mainly occur between functionally adjacent modules, reflecting causal ambiguity in tightly coupled ADS pipelines. Diagnostic inaccuracies often arise when runtime traces confirm the failure but do not isolate the initiating module, especially when an upstream fault is first manifested through Planning or Control behavior.

\begin{answerbox}{RQ1}
  \ourtool{} achieves the best module-level diagnosis by reasoning over system-level causality rather than isolated rule checks. The confusion matrices further show that this gain reflects robust fault attribution across major ADS modules rather than reporting bias toward specific modules.
\end{answerbox}

\subsection{RQ2. Code-level Fault Localization}

\rjupdate{Here we evaluate code-level fault localization within the pre-identified faulty module. By providing each baseline method with the identical annotated module as context, this setup isolates Phase II localization efficacy from Phase I diagnostic accuracy.}
We compare against AutoFL, LocAgent, and Claude Code, which perform LLM-based repository exploration without ADS-specific runtime-intent analysis.

\begin{table}[t]
  \centering
  \caption{Code-Level Fault Localization Accuracy}
  \label{tab:code-fl}
  \small
  \setlength{\tabcolsep}{3.1pt}
  \renewcommand{\arraystretch}{1.10}
  \begin{tabular}{ll|ccc|ccc}
  \toprule
  \multirow{2}{*}{\textbf{Data}} & \multirow{2}{*}{\textbf{Method}} & \multicolumn{3}{c|}{\textbf{Class (\%)}} & \multicolumn{3}{c}{\textbf{Func. (\%)}} \\
   &  & \textbf{@1} & \textbf{@3} & \textbf{@5} & \textbf{@3} & \textbf{@5} & \textbf{@10} \\
  \midrule
  \multirow{4}{*}{$D_{inject}$}
  & AutoFL & 17.5 & 17.5 & 23.8 & 14.3 & 17.5 & 39.7 \\
  & LocAgent & 7.9 & 9.5 & 17.5 & 4.8 & 7.9 & 15.9 \\
  & Claude Code & \textbf{57.2} & 57.2 & 66.7 & 41.3 & 49.2 & 73.0 \\
  & \textbf{\ourtool{}} & 46.0 & \textbf{65.1} & \textbf{74.6} & \textbf{42.9} & \textbf{61.9} & \textbf{90.5} \\
  \midrule
  \multirow{4}{*}{$D_{real}$}
  & AutoFL & 11.1 & 11.1 & 33.3 & 0.0 & 11.1 & 22.2 \\
  & LocAgent & 11.1 & 33.3 & 44.4 & 0.0 & 33.3 & 44.4 \\
  & Claude Code & 11.1 & 33.3 & 55.6 & 11.1 & 11.1 & 33.3 \\
  & \textbf{\ourtool{}} & \textbf{55.6} & \textbf{88.9} & \textbf{88.9} & \textbf{33.3} & \textbf{66.7} & \textbf{88.9} \\
  \bottomrule
  \end{tabular}
\end{table}

Table~\ref{tab:code-fl} shows that \ourtool{} performs better at larger Top-$k$ cutoffs, function granularity, and real bugs, with the exception of Claude Code's Class@1 result on $D_{inject}$. This suggests that injected faults can expose obvious class-level cues, while \ourtool{} gives more reliable rankings when runtime and intent evidence are needed. AutoFL and LocAgent lag behind because their search mainly follows repository structure, static signals, or generic LLM navigation, which can surface semantically related code but not the failure-relevant code region.

Claude Code is a stronger generic code agent and retrieves plausible faulty classes on injected faults. Its advantage weakens, however, when the task requires ranking the faulty function, especially on real bugs. This pattern suggests that broad repository exploration is useful for finding candidate classes, but is less reliable for separating the runtime-relevant faulty function from nearby implementation logic. \ourtool{} improves this step by restricting the search to DES-derived runtime-relevant code and ranking candidates through design-implementation inconsistency analysis.

\rjupdate{While function-level misses occur, \ourtool{}’s predictions generally cluster within the identical causal neighborhood. For intricately branched industrial code, adjacent functions on the same execution chain remain highly plausible, even under strict exact-match scoring constraints.}

\begin{answerbox}{RQ2}
  \ourtool{} achieves the strongest overall code-level localization, with clearer gains at function granularity and on real-world bugs. By pruning to runtime-relevant code and comparing intended and implemented behavior, it ranks faulty functions more reliably than generic code-agent exploration.
\end{answerbox}

\subsection{RQ3. Ablation Study
}

To validate the necessity of each design component in \ourtool{}, 
we conduct a comprehensive ablation study on $D_{real}$. 
We evaluate three variants, each removing one of the following components: 
Scenario Abstraction, DES, or Inconsistency Analysis.

\begin{itemize}[leftmargin=*]
  \item \textbf{w/o Scenario Abstraction:} Remove \textit{Scenario Abstraction}; the Hypothesis Agent diagnoses faults directly from numerical traces and STL monitor outputs.
  \item \textbf{w/o DES:} Replace the \textit{DES} with a static call graph, forcing navigation based only on static code dependencies.
  \item \textbf{w/o Inconsistency Analysis:} Replace the dual-stream comparison with a unified RAG process, where the LLM receives specifications and code together and directly scores candidate functions in a single pass.
\end{itemize}

\begin{table}[t]
  \centering
  \caption{Ablation Study on $D_{real}$}
  \label{tab:ablation_combined}
  \begin{threeparttable}
  \small
  \setlength{\tabcolsep}{1.2pt}
  \renewcommand{\arraystretch}{1.06}
  \begin{tabular}{l|ccc|cc|cc}
  \toprule
  \multirow{3}{*}{\textbf{Variant}} & \multicolumn{3}{c|}{\textbf{Phase I}} & \multicolumn{4}{c}{\textbf{Phase II}} \\
  \cmidrule(lr){2-4}\cmidrule(lr){5-8}
   & & & \multicolumn{2}{c|}{\textbf{Class}} & \multicolumn{2}{c}{\textbf{Func.}} \\
   & \textbf{Acc.} & \textbf{Steps} & \textbf{TO} & \textbf{@1} & \textbf{@5} & \textbf{@3} & \textbf{@10} \\
  \midrule
  \rowcolor{gray!10} \textbf{\ourtool{} (Full)} & \textbf{88.9\%} & \textbf{4.0} & \textbf{0.0\%} & \textbf{22.2\%} & \textbf{77.8\%} & \textbf{33.3\%} & \textbf{77.8\%} \\
  \midrule
  \textit{w/o Scen. Abs.} & 33.3\% & 15.0+ & 50.0\% & \multicolumn{4}{c}{\textit{N/A$^\dagger$}} \\
  \midrule
  \textit{w/o DES}$^*$ & -- & -- & -- & 0.0\% & 22.2\% & 0.0\% & 22.2\% \\
  \textit{w/o Incon. Anal.}$^*$ & -- & -- & -- & 0.0\% & 33.3\% & 11.1\% & 33.3\% \\
  \bottomrule
  \end{tabular}
  \vspace{2pt}
  \begin{tablenotes}[flushleft]
    \footnotesize
    \item \textit{TO}: Timeout; \textit{Scen. Abs.}: Scenario Abstraction; \textit{DES}: Dynamic Execution Skeleton; \textit{Incon. Anal.}: Inconsistency Analysis.
    \item[] $^*$Phase II uses the ground-truth module. $^\dagger$Phase II is skipped when Phase I fails.
  \end{tablenotes}
  \end{threeparttable}
\end{table}

\textit{Impact of Scenario Abstraction.}
As shown in Table~\ref{tab:ablation_combined}, removing \textit{Scenario Abstraction} substantially harms both efficiency and accuracy: the average interaction length increases from 4.0 to over 15.0 steps with a 50.0\% timeout rate, while Phase I accuracy drops from 88.9\% to 33.3\%. 
This indicates that relying solely on numerical traces and monitor outputs, without a holistic scene description, is insufficient for accurate module diagnosis and also increases the complexity of the analysis.

\textit{Impact of the DES}. 
The \textit{w/o DES} variant uses a static dependency graph instead of the runtime-specific skeleton, causing a sharp drop in localization accuracy (0.0\% Acc@1). In polymorphic frameworks such as Apollo, static analysis leaves many irrelevant subclass implementations in the search space because it cannot resolve runtime dispatch. DES avoids this by pruning the search to actually executed logic.

\textit{Impact of Inconsistency Analysis}. 
The \textit{w/o Inconsistency Analysis} variant also shows a clear accuracy drop. Without explicit target-implementation comparison, the LLM tends to favor candidates that look plausible in the retrieved context, even when no concrete semantic mismatch is identified. The full version avoids this by scoring structured inconsistency signals and calibrating them with design- and code-side reliability, which yields more stable rankings and removes many false positives.

\begin{answerbox}{RQ3}
  Ablation results confirm the necessity of all three core components for superior diagnostic accuracy. Removing any of them noticeably weakens \ourtool{} in either accuracy or efficiency.
\end{answerbox}
\subsection{RQ4: End-to-End Performance}

This research question evaluates the practical utility of \ourtool{} as a complete diagnostic pipeline—from observing a system-level failure to producing an actionable fault localization report—without
any intermediate ground truth (e.g., pre-identified faulty modules).

\begin{table}[t]
  \centering
  \caption{End-to-End Fault Localization Performance}
  \label{tab:e2e-performance}
  \small
  \setlength{\tabcolsep}{3.2pt}
  \renewcommand{\arraystretch}{1.08}
  \begin{tabular}{llccc}
  \toprule
  \textbf{Data} & \textbf{Method} & \textbf{Class@1} & \textbf{Class@5} & \textbf{Func.@10} \\
  \midrule
  \multirow{3}{*}{$D_{inject}$}
    & Claude Code & 25.4\% & 28.6\% & 25.4\% \\
    & Claude Code + Skills & 39.6\% & 55.6\% & 63.5\% \\
    & \textbf{\ourtool{}} & \textbf{42.9\%} & \textbf{66.7\%} & \textbf{81.0\%} \\
  \midrule
  \multirow{3}{*}{$D_{real}$}
    & Claude Code & 33.3\% & 55.6\% & 55.6\% \\
    & Claude Code + Skills & 33.3\% & 44.4\% & 55.6\% \\
    & \textbf{\ourtool{}} & \textbf{44.4\%} & \textbf{77.8\%} & \textbf{77.8\%} \\
  \bottomrule
  \end{tabular}
\end{table}

\textit{End-to-End Quantitative Results}.  Results in Table~\ref{tab:e2e-performance} show that Claude Code alone remains limited in the end-to-end setting. Providing our Skills improves its results on $D_{inject}$, suggesting that these scripts provide useful runtime-grounded evidence, but the improvement does not transfer to $D_{real}$. This indicates that the Skills are not sufficient as stand-alone prompts or scripts. \ourtool{} performs best because it selects and validates hypotheses, restricts code search through DES, and ranks candidates by intent inconsistency.

On $D_{real}$, Claude Code is slightly stronger in the end-to-end setting than in isolated Phase II (Table~\ref{tab:code-fl}) for some cutoffs. This likely comes from the additional scenario evidence in the trace and STL report: the annotated faulty module used in isolated Phase II gives module context, but it does not expose the full failure behavior. The remaining gap to \ourtool{} suggests that HINT's gain comes from combining runtime evidence with a structured pipeline for hypothesis validation, DES-guided code restriction, and intent-based ranking.

\textit{Robustness Across LLM Backbones}.
We further evaluate \ourtool{} with four representative backbones from three model families: Claude-series models~\cite{anthropic_claude} (Haiku 4.5 and Sonnet 4.5), GPT-5.4~\cite{openai2024chatgpt}, and GLM-5~\cite{glm52026}. Sonnet 4.5 is the default backbone for reasoning-heavy steps in our prototype, as it provides the best accuracy among the evaluated LLMs while remaining a practical choice for repeated agent calls. Haiku 4.5 represents a lighter Claude-series model, while GPT-5.4 and GLM-5 test whether the pipeline transfers beyond one provider family. In Table~\ref{tab:backbone}, MFR (Mean First Rank) denotes the average rank of the first correct fault location, where lower is better. The results show that class-level localization remains stable across backbones, indicating that Phase I and coarse code retrieval do not depend on a single model. Function-level ranking is more sensitive to backbone quality, with Sonnet 4.5 performing best. Nevertheless,  other backbones still keep the correct fault within practical Top-$k$ inspection ranges. This implies that the structured two-phase pipeline, rather than a single proprietary model, accounts for most of \ourtool{}'s gain.

\begin{table}[t]
  \centering
  \caption{Robustness Across Different LLM Backbones on $D_{real}$}
  \label{tab:backbone}
  \small
  \setlength{\tabcolsep}{4pt}
  \begin{tabular}{l|cc|cc}
  \toprule
  \textbf{Backbone} & \textbf{Class@5} & \textbf{Func@5} & \textbf{Class MFR} & \textbf{Func MFR} \\
  \midrule
  Haiku 4.5  & 66.7\% & 55.6\% & 6.89 & 9.56 \\
  Sonnet 4.5  & {\bf 77.8\%} & {\bf 66.7\%} & {\bf 5.56} & {\bf 7.33} \\
  GPT-5.4     & \textbf{77.8}\% & 44.4\% & 6.22 & 9.44 \\
  GLM-5       & 66.7\% & 55.6\% & 7.00 & 10.67 \\
  \bottomrule
  \end{tabular}
\end{table}

\begin{answerbox}{RQ4}
  \ourtool{} remains effective end-to-end across LLM backbones, showing that its gains come from the structured pipeline rather than a specific model or artifact access alone.
\end{answerbox}

\section{Related Works}

\noindent{\bf ADS testing \& Module Error diagnosis}. ADS testing mainly targets failure discovery through scenario generation, specification checking, and simulation-based exploration~\cite{cheng2023behavexplor,zhou2023specification,sun2022lawbreaker,wang2025moditector,huai2023doppelganger,zhong2022neural,li2020av,kim2022drivefuzz,yan2025demand,guo2024sovar,li2024viohawk,li2023simulation,li2025comprehensive}, while fault diagnosis remains less explored. Existing methods are limited by scope or simulation cost: Diavio~\cite{lu2024diavio} identifies responsible agents, MoDitector~\cite{wang2025moditector} uses module-specific oracles, ACAV~\cite{sun2024acav} and ROCAS~\cite{feng2024rocas} infer events or modules from trace discrepancies under restrictive assumptions, and counterfactual methods such as ADSDx~\cite{huangadsdx} and RVPLAYER~\cite{choi2022rvplayer} require iterative re-simulation and ground-truth substitution. In contrast, \ourtool{} diagnoses failures directly from runtime traces through hypothesis-driven reasoning, avoiding re-execution and extending beyond collision-centric failures.

\noindent{\bf Code Fault Localization}. 
Traditional fault localization includes spectrum-, mutation-, learning-, and slice-based methods~\cite{wong2016survey,abreu2007accuracy,chekam2016assessing,dusia2016recent,demillo1996critical}, while recent LLM-based approaches such as SoapFL~\cite{qin2025s} and LocAgent~\cite{chen2025locagent} perform agentic repository navigation. These methods usually assume software-level failure signals, such as failing tests, stack traces, or coverage spectra. In contrast, ADS failures are typically captured as runtime traces and STL violations, which identify abnormal behavior but do not directly expose the faulty code region. CPS-oriented methods~\cite{kim2020control,bartocci2018localizing} use runtime traces and temporal logic but remain coarse-grained. ADS-specific efforts such as ARIEL~\cite{abdessalem2020automated} and ROCAS~\cite{feng2024rocas} are restricted to simplified settings or specific misconfigurations and still require substantial human interpretation. \ourtool{} bridges this gap by connecting system-level ADS diagnosis with fine-grained code localization in industrial-scale ADS codebases.

\section{Threats to Validity}

\noindent\textbf{Internal Validity.}
\ourtool{} depends on the reasoning quality of its LLM backbone and diagnostic Skills. We mitigate this threat with structured prompting and constrained workflows, and our results remain stable across LLM backbones. 
\rjupdate{The Hypothesis Agent's skills are currently tailored to the specific ADS platform, meaning their quality could potentially impact accuracy and reproducibility when migrating to other platforms. Furthermore, \ourtool{} operates under the assumption of a single dominant faulty module per failure. While this encapsulates a large portion of real-world bugs, it may oversimplify complex multi-module cascades or obscure neighboring-function faults along the shared causal chain.}

\noindent\textbf{External Validity.}
\ourtool{} targets modular ADS with accessible runtime records, documentation, and source code, making end-to-end neural driving systems a different setting. Incomplete artifacts may weaken design- or implementation-side evidence. Our Apollo-based prototype  \rjupdate{depends} on its module decomposition, logs, and code organization. 
\rjupdate{The benchmark encompasses Control, Planning, and Prediction module faults, substantiated by co-simulation logs and reproducible failure traces. Expanding its scope to Perception requires reproducible perception failures with
reliable  annotations.}
The real-world subset is smaller because public reproducible Apollo safety/performance bugs are rare. The framework could still be adapted to modular ADS such as Autoware by replacing recording and code-access interfaces; applying it beyond ADS would require domain-specific evidence extraction and consistency checking.

\section{Conclusion}
We presented \ourtool{}, an automated framework for code-level fault localization in modular ADS. \ourtool{} connects system-level failures to source-code locations by combining runtime-trace diagnosis, DES-guided code restriction, and design-implementation inconsistency analysis. Experiments on Apollo with real-world bugs and injected faults show that \ourtool{} improves both module-level diagnosis and fine-grained code localization over existing baselines. The results also show that its gains come from the structured diagnostic pipeline rather than artifact access or a specific LLM backbone alone.

\section*{Data Availability}
Our anonymous replication package is available at~\cite{hint_replication_package}. It includes benchmark metadata, prompts, Skill templates, model settings, per-case outputs, and evaluation scripts.

\bibliographystyle{IEEEtran}
\bibliography{sample-base}

\end{document}